\documentclass[aps,pre,groupedaddress, twocolumn]{revtex4}
\usepackage{amsmath}
\usepackage{color}
\usepackage{graphicx}
\usepackage{verbatim}

\usepackage[normalem]{ulem}	 
\begin{document}


\title{A Wang-Landau method for calculating Renyi entropies \\ in finite-temperature quantum Monte Carlo simulations}


\author{Stephen Inglis}
\email[]{dartonias@gmail.com}
\affiliation{Department of Physics and Astronomy, University of Waterloo, 200 University Avenue, Ontario , Canada, N2L 3G1}
\author{Roger G. Melko}
\affiliation{Department of Physics and Astronomy, University of Waterloo, 200 University Avenue, Ontario , Canada, N2L 3G1}
\affiliation{Perimeter Institute for Theoretical Physics, Waterloo, Ontario N2L 2Y5, Canada}


\date{\today}

\begin{abstract}
We implement a Wang-Landau sampling technique in quantum Monte Carlo (QMC) for the purpose of calculating the Renyi entanglement entropies and associated mutual information.
The algorithm converges an estimate for an analogue to the density of states for Stochastic Series Expansion QMC allowing a direct calculation of Renyi entropies without explicit thermodynamic integration.
We benchmark results for the mutual information on two-dimensional (2D) isotropic and anisotropic Heisenberg models, 2D transverse field Ising model, and 3D Heisenberg model, confirming a critical scaling of the mutual information in cases with a finite-temperature transition.
We discuss the benefits and limitations of  broad sampling techniques compared to standard importance sampling methods.

\end{abstract}


\maketitle

\section{Introduction}

Entanglement is a measure of the ties that bind quantum mechanical particles across space and time.
In a quantum many-body wavefunction, the amount of entanglement between two sub-regions of a system, A and B, is often quantified through a functional weighting of the eigenvalues $\lambda_i$ of the reduced density matrix $\rho_A = {\rm Tr}_B \rho$, where there is always some set of $| \phi_i \rangle$ such that $\rho = \sum_i \lambda_i |\phi_i \rangle \langle \phi_i |$. 
This is used to define the so-called Renyi entanglement entropies \cite{renyi},
\begin{equation}
S_{\alpha}(A)=\frac{1}{1-\alpha} \ln \left[{ {\rm Tr}(\rho_A^\alpha) }\right], \label{Renyi}
\end{equation}
where the limit $\alpha=1$ is the familiar von Neumann entanglement entropy.
The scaling of $S_1$ in particular has been used to study and classify the groundstate phases and phase transitions of condensed matter systems \cite{Holzhey,VidalC,Fradkin,KP,LW, Max}.
Until recently it could only be accessed in numerical simulations through complete knowledge of the groundstate wavefunction -- restricting its study to Hamiltonians which can be solved by exact diagonalization or related methods (e.g. density matrix renormalization group) \cite{white_dmrg}.

In 2010, Hastings {\it et.~al.}~introduced a procedure to calculate Renyi entropies for integer $\alpha \ge 2$ in zero-temperature projector quantum Monte Carlo (QMC) simulations through the expectation value of a {\it swap} operator \cite{swap}.
This operator acts on two independent copies of the QMC-sampled configuration, literally ``swapping'' basis states in region A between the two copies.
Since introduction of the algorithm, there has been a flurry of activity in calculating Renyi entropies numerically using {\it swap}-related techniques \cite{grover1,grover2,grover3,mcminis1,mcminis2}.
Soon thereafter \cite{roger_multisheet_sse}, it also was demonstrated that Renyi entropies could  be calculated in finite-temperature QMC through adaptation of a well known ``replica'' trick \cite{calabrese_c,drum}, whereby the $d+1$ dimensional QMC simulation cell is doubled (in the case of $S_2$) in size in its imaginary time direction \cite{replicaMC1,Buividovich}.
The Renyi entropy is then related to the logarithm of the ratio of the partition function of the replicated system ($Z[A,2,T]$) to the partition function of the original system, $Z(T)$.
Unlike the {\it swap} method that gives a Renyi entropy directly at $T=0$, the replica trick calculated for two simulations (one for $Z[A,2,T_0]$ and one for $Z(T_0)$) alone does {\it not} give the Renyi entropy at the temperature $T_0$.
Typically, a careful integration over many simulations from $T = \infty$ to $T = T_0$ must be performed to calculate the Renyi entropy \cite{roger_multisheet_sse,roger_MI}.
Other approaches are possible, including a recent idea that allows direct calculation of the ratio of two partition functions at a fixed temperature without integration \cite{roscilde_qmc}, which effectively implements ideas of the \emph{swap} operator at finite temperature.
Ultimately, such ideas should be wed into the application of broad histogram techniques, which are natural solutions for Monte Carlo simulations which require calculating the partition function.

In this paper, we use an advanced Wang-Landau \cite{WL} sampling technique for calculating the Renyi entropy in finite-temperature QMC that avoids the need for many separate simulations, and allows access to the Renyi entropy at any arbitrary temperature between $T=\infty$ and some chosen cut-off temperature.
We outline the technique for the specific case of $S_2$ using a Stochastic Series Expansion (SSE) QMC.
We then demonstrate the method by looking at the finite-size scaling of the Renyi mutual information in three variants of the spin-1/2 XXZ model in two and three dimensions and the 2D transverse field Ising model.


\section{Methods}
In this section we discuss the specific implementation details of the Wang-Landau method in the context of QMC and calculating the mutual information between two sub-regions A and B.
This particular implementation is guided by the established SSE QMC \cite{sandvik_sse2,sandvik_sse,Sengupta} and recent extensions using a modified simulation topology \cite{roger_multisheet_sse}.
For adding Wang-Landau sampling to the SSE QMC we find the method outlined by Troyer, Wessel and Alet (2003) \cite{troyer_WL_sse} is a simple and concise starting point, although it has similar limitations that have been found also to exist in classical Wang-Landau \cite{WL_1overt_fixing}.
Going beyond Wang-Landau approaches to more general broad histogram sampling, we implement a more advanced technique  detailed by Wessel, Stoop, Gull, Trebst, and Troyer (2007) \cite{Wessel_MRT} that addresses some of the shortcomings with the original Wang-Landau approach, but still provides the density of states needed to calculate thermodynamic properties (including the Renyi entropy) over a large range of temperatures.

\subsection{Stochastic series expansion}
The SSE \cite{sandvik_observables,sandvik_sse} is a framework similar conceptually to world line methods \cite{scalapino_worldline,loop1,Evertz} in which one simulates a $d$-dimensional quantum system using a $(d+1)$-dimensional classical simulation.
To begin, we recall the standard SSE implementation, which represents the partition function as
\begin{align}
Z =  \text{Tr}\Bigl[ e^{-\beta H} \Bigr] 
=  \sum_{\psi} \bigl< \psi \bigr| \sum_{n=0}^{\infty} \frac{(-\beta)^n H^n}{n!} \bigl| \psi \bigr> ,
\end{align}
where $\sum_\psi$ represents the trace over some complete basis of states for the system.
We rewrite the above by collecting all the prefactors to the left and inserting the resolution of the identity between each power of the Hamiltonian.
Since the Hamiltonian is a sum, we can group operators into pieces that act on certain groups of sites and by whether the are diagonal or off diagonal in our basis of choice.
These grouped pieces of the Hamiltonian we refer to as ``bond'' operators $H_{a,b}$ where $a$ differentiates diagonal and off diagonal terms and $b$ denotes a grouping of sites.
For our formalism, $a=1$ represents diagonal terms while $a=2$ represents off diagonal terms, while $b$ specifies the bond connecting two sites that the Hamiltonian piece operates on.
Using this formalism we obtain the SSE representation of the partition function,
\begin{align}
Z = & \sum_n \frac{(-\beta)^n}{n!}  \sum_{\{\psi\}} \sum_{i=0}^n \bigl< \psi_i \bigr| H_{a_i,b_i} \bigl| \psi_{i+1} \bigr> , \label{eq:Z_full}
\end{align}
where $\sum_{\{\psi\}}$ represents the sum over all valid configurations $\{\psi_0 \ldots \psi_{n+1}\}$ with the condition $\bigl< \psi_0 \bigr| = \bigl< \psi_{n+1} \bigr|$.
If we allow the sum to go from zero to infinity and include all possible lists of operators $H_{a_i,b_i}$ for each $n$, this formulation is exact.
In practice, contributions above some large value $N$ are numerically insignificant and can be ignored.
The cutoff $N$ is always larger than any visited $n$ in the importance sampling SSE QMC, and as such does not affect the statistics except to allow us to represent the set of operators as a finite list.

The two updates we use to efficiently sample only non-zero weights are referred to as the {\it diagonal update} and the {\it directed loop update} \cite{sandvik_sse}.
In the diagonal update we insert and remove diagonal operators from the list, changing $n$ and the total weight, but not changing $\bigl< \psi_0 \bigr|$.
We use the change in weight at each insertion or removal and the usual Boltzmann condition to accept or reject these proposed moves.
In the directed loop updates we connect all the Hamiltonian elements by what sites they interact with at each insertion of the identity in the expanded trace $\bigl< \psi_i \bigr|$.
We then allow a loop to traverse this linked network of sites and change the operators (and spin configurations) it passes through, changing the \emph{type} of each operator (and hence total weight) but not changing \emph{which} sites each operator interacts with.
Using these two updates our simulation is able to satisfy both ergodicity and detailed balance.

\begin{figure}
\includegraphics[width=0.75\columnwidth]{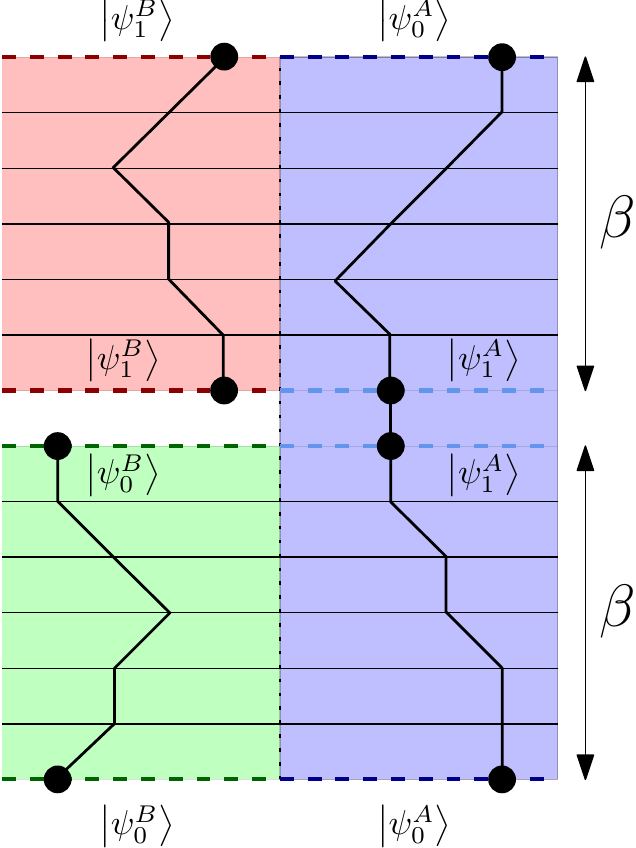}
\caption{Boundary conditions of the modified simulation, related to Equation~\ref{eq:mod_bc}.
States with the same labels represent where there periodic boundary conditions of the simulation cell are, while lines represent the paths of up spins (or bosons, depending on context) in the imaginary time direction.
\label{fig:modified_bc}
}
\end{figure}

Recently, the SSE QMC was adapted to allow for the calculation of Renyi entropy between two regions A and its complement B \cite{roger_MI}.
To do this, the so-called ``replica trick''\cite{roger_replica} was employed to estimate Eq.~(\ref{Renyi}):
\begin{align}
\text{Tr}(\rho^{\alpha}_A) =& \frac{Z(A,\alpha,\beta)}{Z(\beta)^{\alpha}} ,
\end{align}
allowing the calculation the Renyi entropies through:
\begin{align}
S_\alpha(\rho_A) =& \frac{1}{1-\alpha} \bigl( \ln[Z(A,\alpha,\beta)] - \alpha \ln[Z(\beta)] \bigr) ,  \label{FreeEdiff}
\end{align}
Here, the two partition functions represent two separate QMC simulations: $Z(\beta)$ is the ``usual'' partition function for the physical
system under study, and $Z(A,\alpha,\beta)$ is a replicated partition function of the form,
\begin{align}
Z(A,\alpha,T) =  \sum_{{\alpha \beta}} &
\langle \psi^A_0 | \langle \psi^B_0 | e^{-\beta H} |\psi^A_1 \rangle |\psi^B_0 \rangle \nonumber \\
&\otimes \langle \psi^A_1 | \langle \psi^B_1 | e^{-\beta H} |\psi^A_0 \rangle |\psi^B_1 \rangle,
\label{eq:mod_bc}
\end{align}
where $\langle \psi^A_i |$ represents the basis containing spins in A, while $\langle \psi^B_i |$ represent the spins not in region A and, as illustrated in Fig.~\ref{fig:modified_bc}, the boundary of region A is ``stitched'' together between the two replicas in the imaginary
time direction.

It is apparent from Eq.~(\ref{FreeEdiff}) that the calculation of Renyi entropies involves a procedure very similar to calculating a difference of free energies.  
As is well know, neither the free energy, $\ln[Z(T)]$, nor in fact the partition function itself is accessible directly through importance sampling.
Previous calculations of $S_{\alpha}$ were obtained using a thermodynamic integration of internal energies $E$, obtained over many separate simulations (or one ``annealing'' simulation)\cite{roger_multisheet_sse,roger_MI} or by simulations that sample the ratio of two distinct partition functions at a fixed temperature \cite{roscilde_qmc}.
However, techniques involving extended ensembles have been employed for decades in the calculation of free energies and related quantities in a variety of Monte Carlo techniques.
Below, we discuss the adaptation of one of these, Wang-Landau sampling \cite{WL}, to the calculation of Renyi entropies in finite-temperature QMC techniques.

\subsection{Wang-Landu sampling}
In contrast to importance sampling where we sample states using the thermal weights of the system and take an unweighted average of observables, Wang-Landau sampling seeks to use the sampling process itself to determine the distribution of an unknown function.
If we consider Eq.~(\ref{eq:Z_full}) the partition function can be re-written as
\begin{align}
Z = & \sum_n \beta^n g(n) , \label{eq:zb}
\end{align}
where $n$ is the number of operators in the SSE formalism, Eq.~(\ref{eq:Z_full}).
Since in the SSE the energy can be calculated using the estimator \cite{sandvik_estimators} $E = -\bigl< n \bigr> / \beta$, by knowing $g(n)$ we are also able to calculate the energy and related variables like the specific heat for all temperatures (assuming complete knowledge of $g(n)$).
In addition, we are able to explicitly calculate the partition function for any temperature (up to some cutoff), something not possible when using importance sampling.

Unlike importance sampling in which we sample states with weights derived from the Boltzmann distribution, the Wang-Landau method samples using the weighting $W = g^{-1}(E)$, (or $g^{-1}(n)$ in the quantum case) ensuring that all energies are sampled with equal frequency.
Since we do not know $g(E)$ {\it a priori} we start with an estimated density of states, typically flat, and sample the system assuming that our estimate is good.
If $g(E)$ is correct for all energy (up to a scaling factor) then the histogram of energies sampled will be flat---if the histogram is larger for some energy $E'$, than our estimate for $g(E')$ is too low for that energy, and vice versa.
In this way we use the sampling histogram to iteratively update the density of states until the histogram is flat and $g(E)$ is converged.

\subsection{Quantum Monte Carlo and Wang-Landau}
In this section we will describe all the changes added to the SSE QMC necessary for performing quantum Wang-Landau.
As mentioned before, previous work \cite{troyer_WL_sse} has successfully implemented Wang-Landau sampling in SSE QMC, as well as extensions of the method \cite{Wessel_MRT}.
Our implementation mainly focuses on using Wang-Landau (as a proof of concept) and includes recent advances made in the classical Wang-Landau formalism \cite{WL_1overt_fixing}.

The first challenge with all Wang-Landau techniques is representing the sampling function, here $g(n)$, in a tractable way.
This problem of internal representation is that $g(n)$ ranges over $2000$ orders of magnitude (for larger systems of typical Hamiltonians).
Given computer precision, the common solution that works here is to only keep track of the natural logarithm of $g(n)$.
This representation is entirely compatible with the simulation, and allows for the largest amount of precision without using special data types.
For the purpose of the below discussion, we will let $G(n) =\ln(g(n))$.

The biggest change from the importance sampling simulation is a modification of the diagonal update \cite{troyer_WL_sse} in which the number of operators (the $n$ in $g(n)$) is changed.
In the normal simulation, we compare the the weight before and after, giving us the probability of adding an operator as
\begin{align}
P = & \min \biggl\{ \frac{\bigl< \psi_i \bigr| H_{1,b} \bigl| \psi_{i+1} \bigr> \beta}{M-n} , 1 \biggr\},
\end{align}
where $H_{1,b}$ is a diagonal operator acting on bond $b$ and $n$ is the number of operators.
When using Wang-Landau this update changes to
\begin{align}
P = & \min \biggl\{ \frac{\bigl< \psi_i \bigr| H_{1,b} \bigl| \psi_{i+1} \bigr> e^{G(n)-G(n+1)}}{M-n} , 1 \biggr\},
\end{align}
using the current internal estimate of $G(n)$.
Then, whether we accept or reject the move, we update
\begin{align}
g(n) = & g_{\text{old}}(n) \times f ,\\
G(n) = & G_{\text{old}}(n) + F ,
\end{align}
using the current $n$ after the update.
$F=\log(f)$ is the refinement parameter used to converge the function $G(n)$.
The refinement parameter starts as some large value (in our case $e$, the natural log) and when the current histogram is found to be flat ($\pm 10$ percent of the mean for each bin) $F$ is reduced by
\begin{align}
F = F_{old}/2,
\end{align}
and the histogram is reset.
The directed loop update is carried out as before \cite{sandvik_sse2} (after each diagonal update), but $G(n)$ is not changed or used in this update since the number of operators does not change.

The above algorithm works well when the estimate of $G(n)$ is far from convergence, but more recent work on the classical Wang-Landau algorithm \cite{WL_1overt_fixing} has shown that a small saturated error may not converge with this method.
The concept is simple: when the system is close to converged, say the maximum deviation of $G(n)$ from the true function is less than our flatness tolerance, then the simulation will assume the histogram is flat (within tolerance) after a small amount of sampling.
This results in the refinement factor getting small very quickly without changing $G(n)$ significantly, causing early noise in the sampling of $G(n)$ to become frozen in.
To correct for this, we keep track of the number of full diagonal updates (update over all elements of the operator list) and directed loop updates as a number $t$.
We use the aforementioned update until we find that $F < 1/t$, after which we ignore the histograms and instead set the refinement parameter to
\begin{align}
F = 1/t.
\end{align}
This slower refinement is proven to converge the function $G(n)$.
This choice of $F$ ensures both that $F$ is decreasing, and that any saturated error can be overcome through enough simulation, as $\sum_{n=t}^{N_{MC}} 1/n > C$ can be satisfied for any $C$ (representing the saturated error) and $t$ (representing the current Monte Carlo step) given a large enough $N_{MC}$ (representing the number of total Monte Carlo steps).

\subsection{Ensemble optimization}
If we move to more general broad histogram techniques, a good extension for SSE QMC is the ``minimum round trip'' approach \cite{Wessel_MRT}.
This technique tends to reduce the statistical error (shown in Figure~\ref{fig:g_deriv}) by re-shifting the sampling effort to the areas contributing most to the error, thereby being more efficient overall.

To briefly summarize the technique, it keeps track of separate histograms for upward and downward moving walkers, where the walkers are moving in the one dimensional space of {\em number of operators}, $n$.
A walker is ``upward'' if it most recently visited $n=0$, and ``downward'' walkers are those that most recently visited $n=N$ (where $N$ is our cutoff in the number of operators). 
With these two histograms we have both the total sampling (upward plus downward) plus the rate of change in the fraction of upward (or downward) walkers as a function of $n$.
The concept is, if there is a point in the simulation where the fraction of upward walkers change sharply, this is interpreted as the simulation having trouble tunnelling through a barrier in configuration space.
To pass through such barriers more easily, we weight the simulation to locations where the change in the fraction of walkers is large.
In addition, the sampling is weighted by the inverse number of total samples---this alone would simply be Wang-Landau sampling (in that the weights would converge to the inverse density of states), but the addition of the second condition skews the weights.
This process is iterated until the modified weights converge.

With the modified distribution we can sample and take the product of the histogram of visits and our skewed weights to get the density of states (in Wang-Landau the histogram of visits is flat, and so this product is not necessary in principle), and once we have this we can proceed with an identical analysis to our treatment of generating the density of states with Wang-Landau alone.
We implement this method for the 2D transverse field Ising model and present and discuss the results in Section~\ref{sec:results}.

\subsection{Calculating observables}
Unlike importance sampling where the calculation of observables is aggregated over the course of the simulation, Wang-Landau sampling produces a function from which observables can be calculated after the simulation.
In this section we will discuss the challenges associated with using the function $G(n)$ to calculate observables, and the variety of data that we calculate in relation to entanglement.

The first step in using $G(n)$ is normalizing the function, since Wang-Landau converges $G(a)-G(a+1)$ for all $a$, but not the absolute magnitude of either.
By examining the partition function at infinite temperature we see that
\begin{align}
Z(\beta=0) & = \sum_{n=0}^{\infty} \beta^n g(n) = g(0) ,\\
\ln[Z(\beta=0)] & = S(\beta=0) = G(0) ,
\end{align}
where $S(\beta=0)$ is the entropy at infinite temperature.
In a simulation employing spin-$1/2$ particles, at infinite temperature each spin is independent and the entropy is
\begin{align}
S(\beta=0) =  N_s \ln(2) ,
\end{align}
with $N_s$ as the number of spins.
In the modified simulation with $n$ layers, spins \emph{not} in the region are duplicated $(n-1)$ times, and act as independent degrees of freedom.
In this (more general) case, the entropy is
\begin{align}
S_\alpha(A,\beta=0) =  (N_s + (\alpha-1)(N_s-N_A)) \ln(2) ,\label{eq:entropy1}
\end{align}
where $N_A$ is the number of spins in region A.
Using these formulae we can normalize $G(n)$ for any of the simulations by subtracting a constant amount from all elements of $G(n)$ such that $G(0)$ has the correct value.

In addition, if we know the energy at infinite temperature we can gather information about $G(1)$.
This is because if we look at the calculation of energy it can be written as
\begin{align}
E = & -\frac{\sum_n n e^{G(n)} \beta^{n-1}}{\sum_n e^{G(n)} \beta^{n}} + C,\\
\lim_{\beta \rightarrow 0} E = & -e^{G(1)-G(0)} + C,
\end{align}
where $C$ is the sum of all constants subtracted from the Hamiltonian in the SSE formalism.
This gives us a second check for any model in which the high temperature limit of the energy is known.

Using Eq.~(\ref{eq:zb}) and the normalized $G(n)$ we get the equation for the partition function,
\begin{align}
\ln[Z(\beta)] = \ln \Bigl[ \sum_n \beta^n e^{G(n)}\Bigr] .
\end{align}
Although such a calculation does not pose any problems analytically, computing such a large term using fixed precision data types is a small challenge.
To calculate it, first we use a second function $G'(\beta,n)$ defined as
\begin{align}
G'(\beta,n) = G(n) + n \ln(\beta) .
\end{align}
Using this function the log of the partition function can be written as
\begin{align}
\ln[Z(\beta)] = \ln \Bigl[ \sum_n e^{G'(\beta,n)} \Bigr] .
\end{align}
Finally, we factor the entire equation by $K = \underset{n}{\max} \{G'(\beta,n)\}$ to get
\begin{align}
\ln[Z(\beta)] = \ln[K] + \ln \Bigl[ \sum_n e^{G'(\beta,n)-K} \Bigr] .
\end{align}
By doing this, the series can be re-exponentiated, summed, and the natural logarithm taken of the result without the risk of numerical overflow.
Other calculations, such as the energy, use a similar trick to prevent the very large values of the partition function from causing problems in computation.

The method so far allows calculation of estimators derived from $G(n)$, but gives no estimate on the \emph{quality} of the data.
To determine the method's accuracy (without comparing to known results) we use the fact that even using the $1/t$ optimization, the initial noise of the simulation is only proven to asymptotically converge.
By using multiple independent simulations we can estimate any observable $F[G(n)]$ as
\begin{align}
\left\langle F[G(n)]\right\rangle =& \frac{1}{N} \sum_{i=1}^N F[G_i(n)]  ,\label{eq:goodavg}\\
\Delta \left\langle F[G(n)]\right\rangle =& \frac{1}{\sqrt{N}} \biggl( \Bigl[ \frac{1}{N} \sum_{i=1}^N (F[G_i(n)])^2 \Bigr] - \nonumber \\
& \Bigl[ \frac{1}{N} \sum_{i=1}^N F[G_i(n)] \Bigr]^2 \biggr)^{1/2} ,\label{eq:gooderr}
\end{align}
where $\left\langle \cdots \right\rangle$ represents the statistical average of a quantity over the simulation and $G_i(n)$ represents the function generated from the $i^{\text{th}}$ simulation.
The standard deviation of the mean of $\left\langle F[G(n)]\right\rangle$ is represented by $\Delta \left\langle F[G(n)]\right\rangle$.
We denote the estimator as $\left\langle F[G(n)]\right\rangle$ to distinguish our statistical estimate for the observable to the analytically exact $F[G(n)]$ if we had precise knowledge of $G(n)$.

\begin{figure}
\includegraphics[width=0.85\columnwidth]{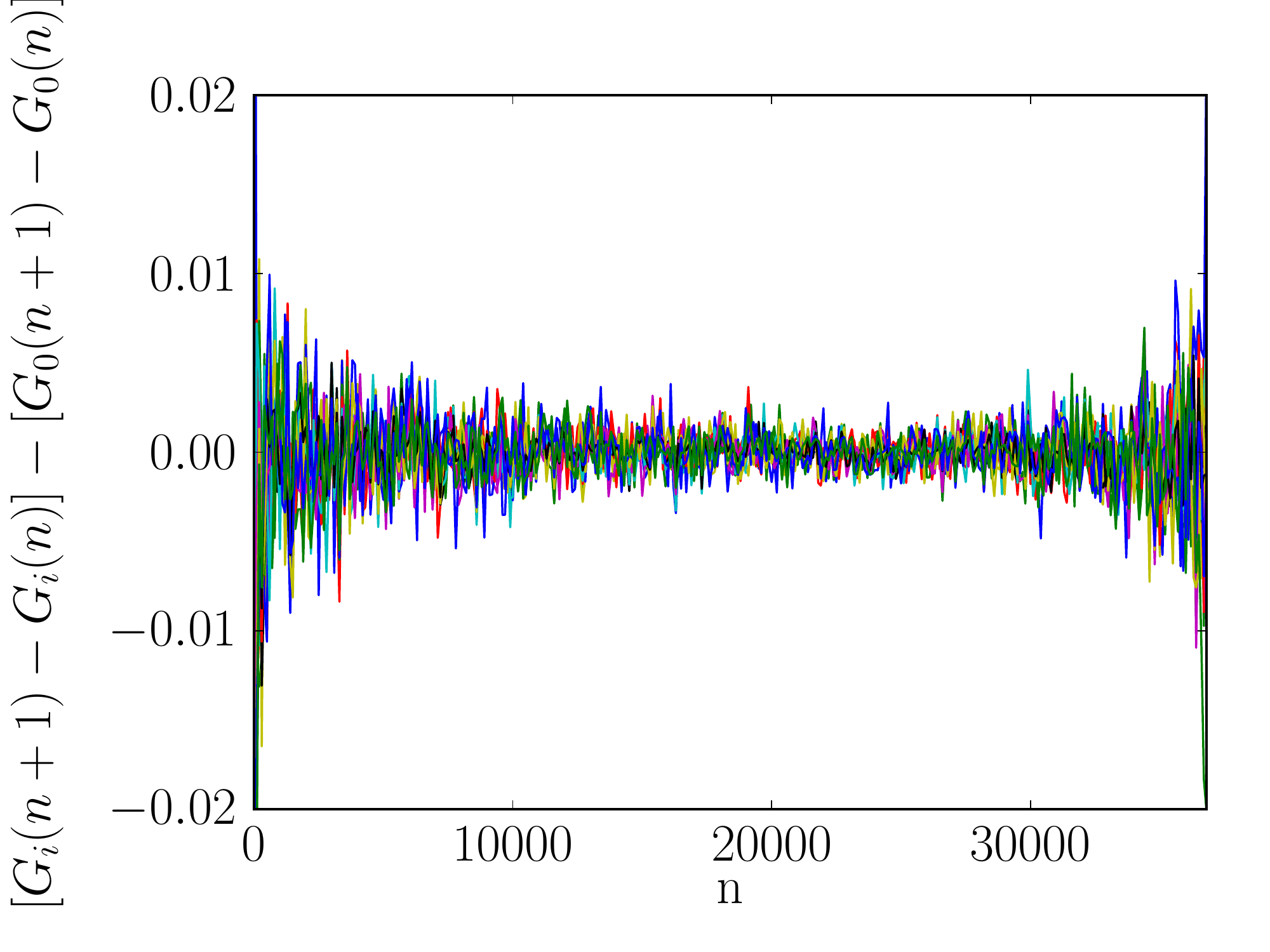}
\caption{Comparison of the first derivative of $G_i(n)$ using $G_0(n)$ (arbitrary) as a baseline for $i={0..15}$.
Each color represents a different independent simulation of the 2D XXZ model of a $32\times 32$ lattice.
The noise is evenly distributed with a larger width for the minimal and maximal values of $n$.
\label{fig:g_deriv}
}
\end{figure}

The noise in $G(n)$ might make it tempting to estimate $F[G(n)]$ as
\begin{align}
\left\langle F[G(n)]\right\rangle =& F[\left\langle G(n)\right\rangle] , \label{eq:badavg}\\ 
\left\langle G(n)\right\rangle =& \frac{1}{N} \sum_{i=1}^N G_i(n) .
\end{align}
We have tested that calculating the estimate of the observables by Eq.~(\ref{eq:goodavg}) gives very similar results to Eq.~(\ref{eq:badavg}), but it does not lend itself easily to calculating the confidence of the the estimate.
Comparison of $G(n)$ from different simulations show that the deviation from the mean is highly correlated---that is to say if $G_i(n)$ is larger than average that $G_i(n+k)$ is also likely to be larger than average for small $k$.
This error is clarified by examining the distribution discrete first derivative of $G_i(n)$ in Fig.~\ref{fig:g_deriv}.
That the first derivative is the quantity that has a random uncorrelated distribution rather than the function itself can be understood by understanding that it is the ratio between adjacent elements, $g(n+1)/g(n)$ or $G(n+1) - G(n)$, that the Wang-Landau simulation converges for all $n$, and through this the algorithm is able to reconstruct $G(n)$.


\section{Results for Mutual Information}
\label{sec:results}
The Wang-Landau technique outlined above is particularly suited for studying finite-temperature properties of the Renyi entanglement entropies.  It's important to note that,
at finite temperature, the Renyi entropies in general no longer obey the property that $S_{\alpha}(\rho_A) = S_{\alpha}(\rho_B)$, since each quantity picks up volume contributions to its scaling from thermal mixing as the temperature is increased from $T=0$.  The analogous quantity that is studied at finite temperatures is therefore the {\it mutual information} (MI), 
\begin{align}
I_\alpha  = S_\alpha (\rho_A) + S_\alpha (\rho_B) - S_\alpha(\rho_{A \cup B}),
\end{align}
which is designed to cancel the volume contributions affecting each Renyi entropy at $T >0$, resulting in a symmetric quantity that reduces to $I_{\alpha} = 2 S_{\alpha}(\rho_A) = 2 S_{\alpha}(\rho_B)$ at $T=0$ assuming $S_\alpha(\rho) = 0$ at zero temperature (which is not true if there is ground state degeneracy).
In QMC, the MI can be calculated for a given region A and its compliment (B) using three simulations---two if $A$ and $B$ are congruent.
In previous work \cite{roger_multisheet_sse} the calculation for the Renyi entropy was done using thermodynamic integration, extending Eq.~(\ref{eq:entropy1}) to finite temperature by integrating energy, similar to how one would calculate thermodynamic entropy in classical systems:
\begin{align}
S_\alpha (\rho_A) = & -S_\alpha (A,\beta=0) + \alpha S(\beta=0)  \nonumber \\
&+ \int_0^\beta \left\langle E \right\rangle_{A,\beta'} d\beta' - \alpha \int_0^\beta \left\langle E \right\rangle_{\beta'} d\beta' ,\label{eq:integrate}
\end{align}
where $\left\langle E \right\rangle_{A,\beta'}$ ($\left\langle E \right\rangle_{\beta'}$) is the energy of the modified simulation cell (normal simulation) at inverse temperature $\beta'$.
Using this integration technique, the MI was studied using QMC in the context of  the anisotropic Heisenberg (or XXZ) model \cite{roger_MI}.
To calculate results using Eq.~(\ref{eq:integrate}) simulations were run at many temperatures to provide the value of $\left\langle E \right\rangle_{A,\beta'}$ over a sufficiently wide range and fine temperature mesh to perform numerical integration.
By studying the finite-size scaling of the $\alpha$-Renyi entropies, the authors discovered that any critical behavior manifest at a finite-temperature phase transition $T_c$ is manifest as approximate crossings of the MI (for different lattice sizes) at $T_c$ and $\alpha T_c$ \cite{roger_MI}.

We first examine the behavior of the MI in the 2D square-lattice XXZ model,
\begin{align}
H = \sum_{\left<ij\right>} \bigl( \Delta S^z_i S^z_j + S^x_i S^x_j + S^y_i S^y_j \bigr) ,
\end{align}
using our Wang-Landau algorithm of the previous section, and compare its performance to the conventional integration technique \cite{roger_MI}.
In addition to the 2D anisotropic Heisenberg model of Ref.~\onlinecite{roger_MI} ($\Delta=4$), we examine the 2D Heisenberg model($\Delta = 1$).
To test an extended Wang-Landau method, we use the 2D transverse field Ising model \cite{sandvik_ising} in the regime where the transverse field is weak and it has a finite temperature phase to an Ising-like ground state.
The Hamiltonian for this model is typically written as
\begin{align}
H = - J \sum_{\left< ij \right>} \sigma^z_i \sigma^z_j - \Gamma \sum_i \sigma^x_i ,
\end{align}
where $\sigma^z$ and $\sigma^x$ are Pauli spin operators with eigenvalues $\pm 1$, and we study the model at $\Gamma = J = 1$.
Finally, we examine the 3D Heisenberg model on a cubic lattice \cite{sandvik_3dheis}, which also has a finite-temperature transition.


\begin{figure}
\includegraphics[width=0.85\columnwidth]{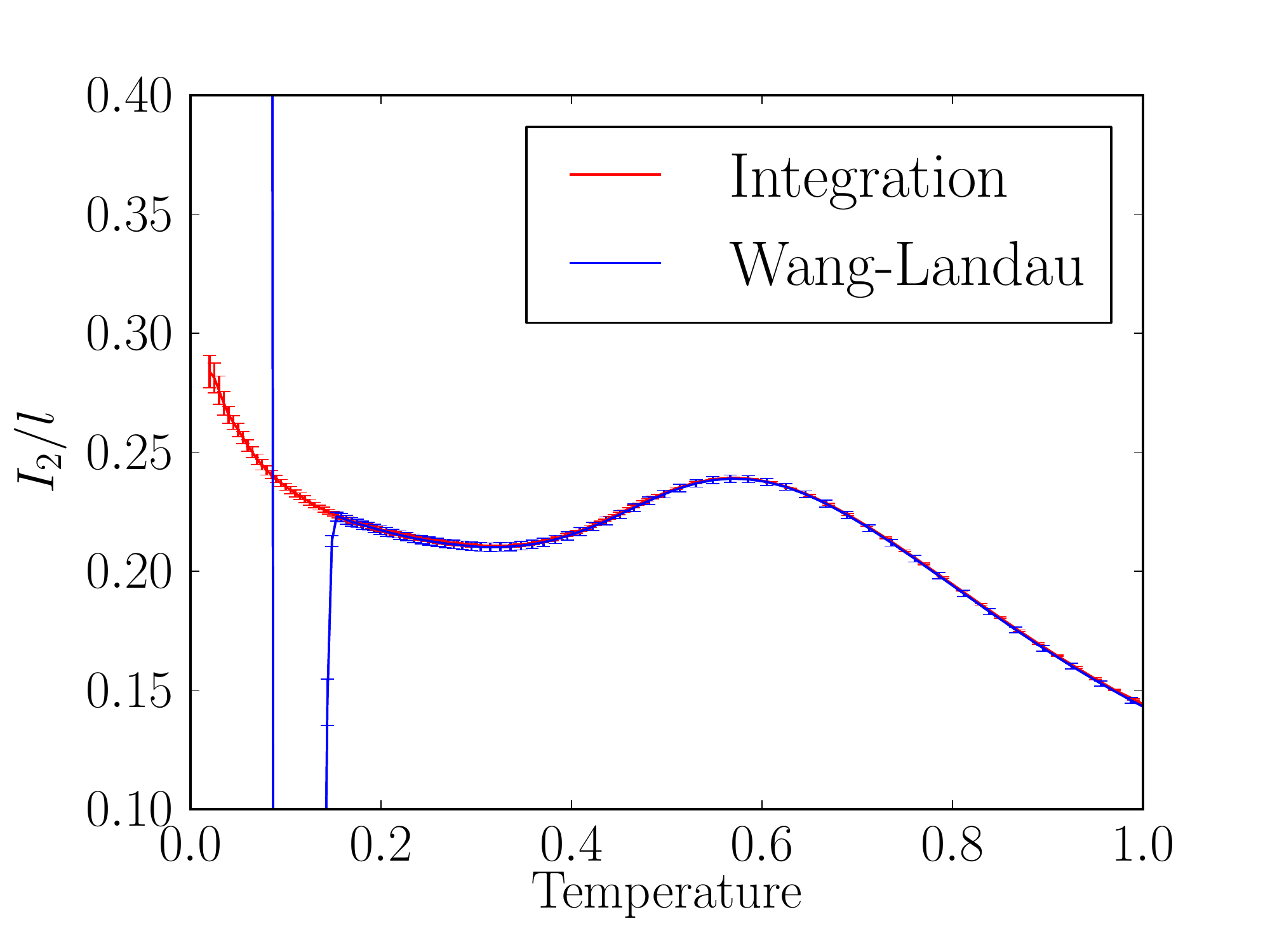}
\caption{$I_2/l$, where $l$ is the boundary length, for the 2D Heisenberg model with $L=12$, comparing the integration of importance sampling and Wang-Landau.
Notice the sharp anomalous behavior in the Wang-Landau for temperatures slightly below the cutoff, here chosen to be $T=0.2$.
This means we can only guarantee results down to that temperature, but not that they necessarily diverge there. \label{fig:wl_cutoff}
}
\end{figure}
To begin, we examine the dependence of the MI on the cutoff error due to our knowledge of $g(n)$ only up to some large (finite) value of $n$.
In the usual SSE QMC the number of operators $n$ sampled is fluctuating, but for a given $\beta$ we can associate an upper bound on the number of operators needed to faithfully represent the system, $N$.
This $N$ can be used to choose an appropriate cutoff in the Wang-Landau algorithm.
Choosing to have a hard cutoff in $n$ corresponds to a cutoff in reliable data at some $\beta$.
This can be chosen to be slightly larger than the $N$ identified for the cutoff $\beta$ in order to ensure that data is still accurate at the $\beta$ of interest---typically 1.2 to 1.4 times $N$.
Results generated for temperatures above the cutoff are accurate, but those below the cutoff exhibit anomalous behavior, as effectively $g(n) = 0$ for $n>N$ (or equivalently, $G(n) = -\infty$).
The result of attempting to generate results below this cutoff is shown in Fig.~\ref{fig:wl_cutoff}.
$G(n)$ is a smooth function for large enough $n$, so there is room for methods which generate data using analytic extensions of $G(n)$ to reduce the anomalous behavior near the lower temperature bound, but in general exactly knowing such an analytic form (for all $n$) would be equivalent to having an analytical expression for the energy as a function of temperature.

We now briefly examine the finite-size scaling of the MI for several examples of $\Delta$ in the XXZ model.
To begin, recall that the anisotropic Heisenberg model (with $\Delta=4$) undergoes a phase transition to a z-axis polarized antiferromagnetic ground state as the temperature is lowered.
The transition temperature was determined previously using the fourth order binder cumulant of the staggered magnetization, finding $T_c = 2.42$ \cite{roger_MI}.
\begin{figure}
\includegraphics[width=0.9\columnwidth]{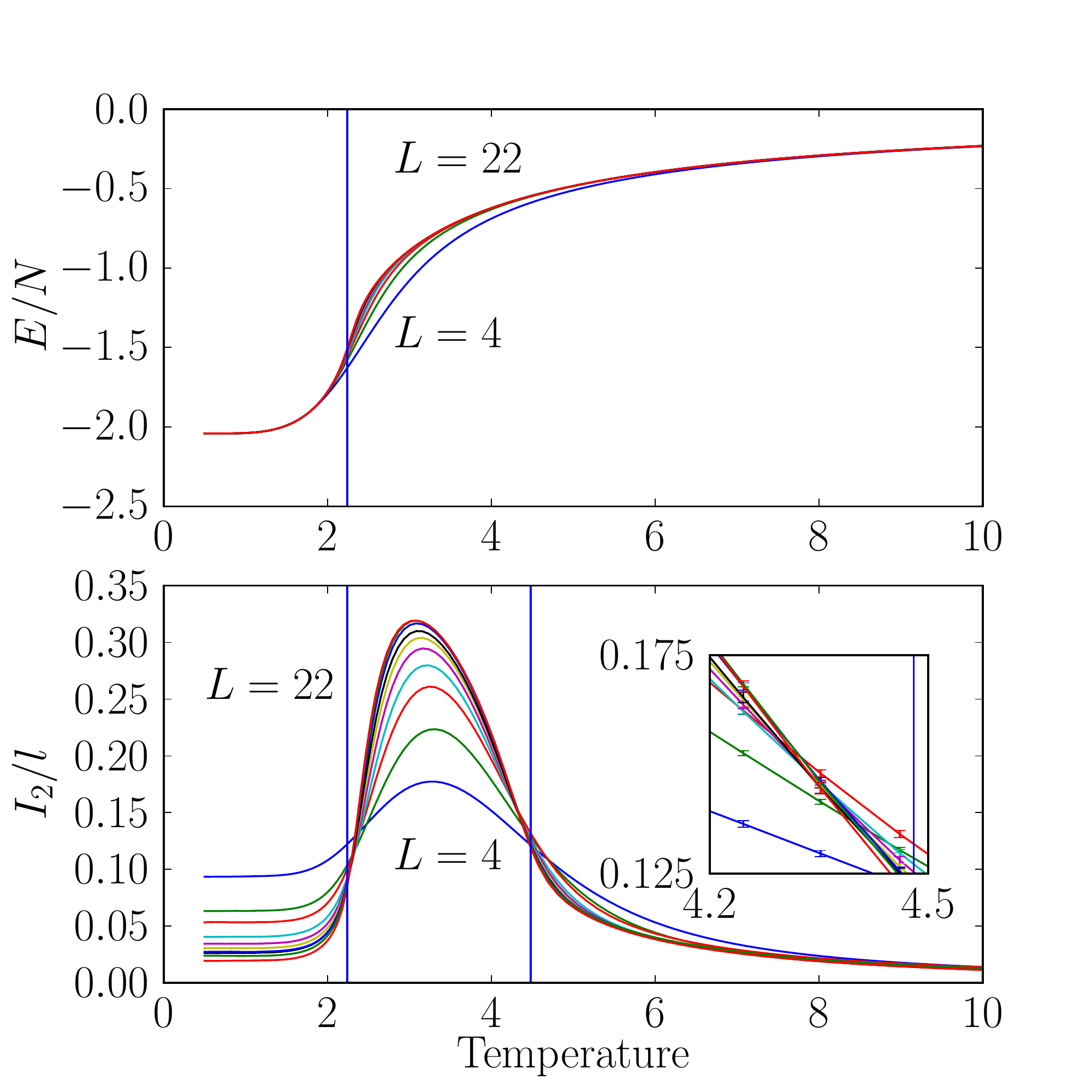}
\caption{The energy per spin and mutual information divided by boundary length for the 2D XXZ model with $\Delta=4$.
Notice the crossings of the mutual information very close to $T_c$ and $2T_c$, marked with vertical lines.
Different lines correspond to lattices are of size $4,6,8,10,12,14,16,18,20$ and $22$ (from lowest to highest at peak).
\label{fig:2d_delta4}
}
\end{figure}
The energy per spin and mutual information $I_2$ divided by boundary size $l=2L$ is shown in Fig.~\ref{fig:2d_delta4}.
The mutual information per system size for different $L$ exhibits a universal scaling \cite{roger_MI}
manifest as approximate
crossings at $T_c$, and $2T_c$ (when using $I_2$ from the second Renyi entropy).
The inset in Figure~\ref{fig:2d_delta4} shows a closeup of the crossing, with the uncertainty of the data points calculated from Eq.~(\ref{eq:gooderr}).

\begin{figure}
\includegraphics[width=0.85\columnwidth]{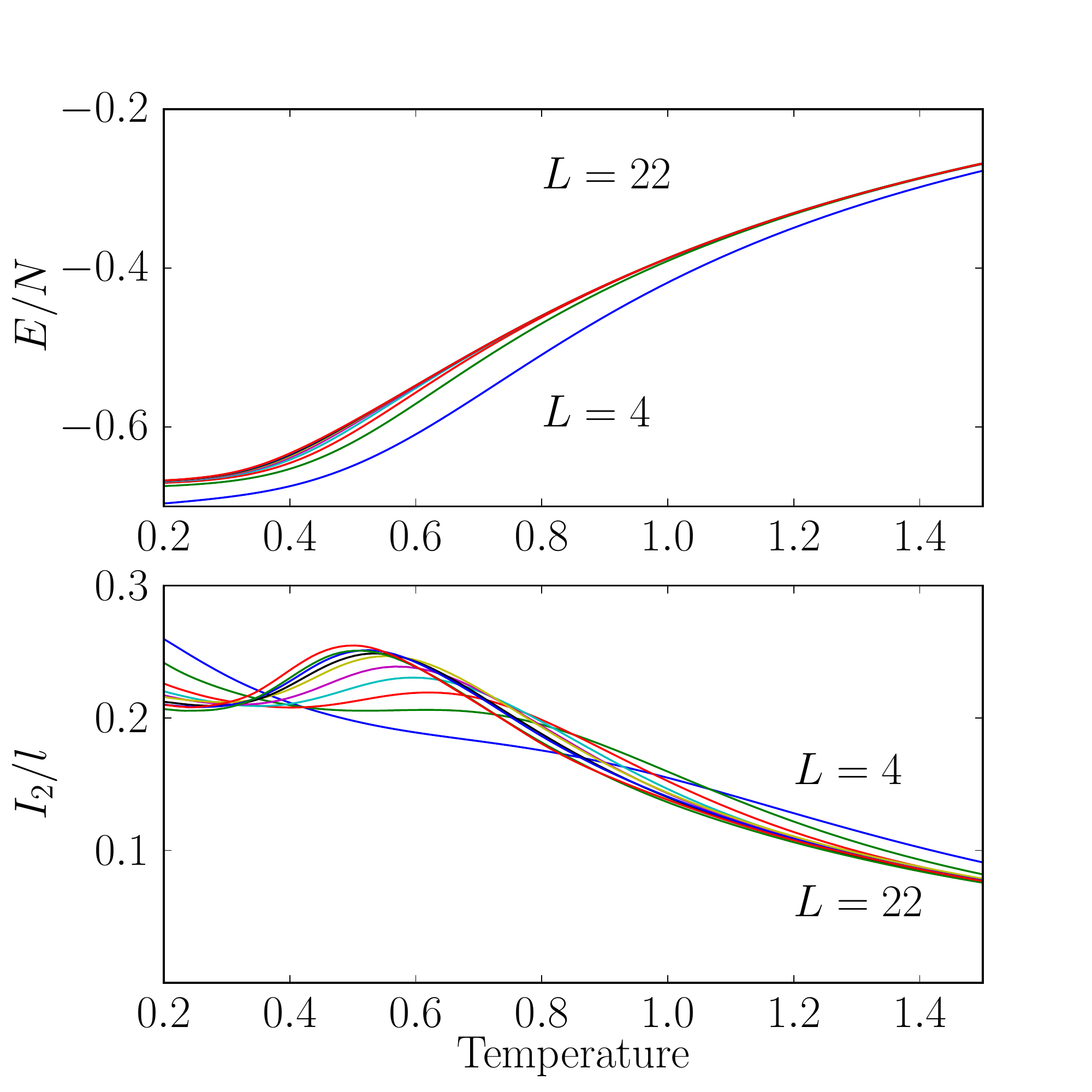}
\caption{The energy per spin and mutual information divided by system size for the 2D Heisenberg model.
This model shows no fixed crossing in $I_2/l$.
Neighboring sizes cross at progressively lower temperatures.
Different lines correspond to lattices are of size $4,6,8,10,12,14,16,18,20$ and $22$ (from lowest to highest central peak).
\label{fig:2d_heis}
}
\end{figure}
It is well known that the 2D isotropic Heisenberg model ($\Delta=1$) has no finite temperature phase transition.  When examining the mutual information, we do not see any fixed crossing in the curves as a function of $L$, as shown in Fig.~\ref{fig:2d_heis}; rather, the crossings move towards $T=0$ for increasing $L$.

\begin{figure}
\includegraphics[width=0.85\columnwidth]{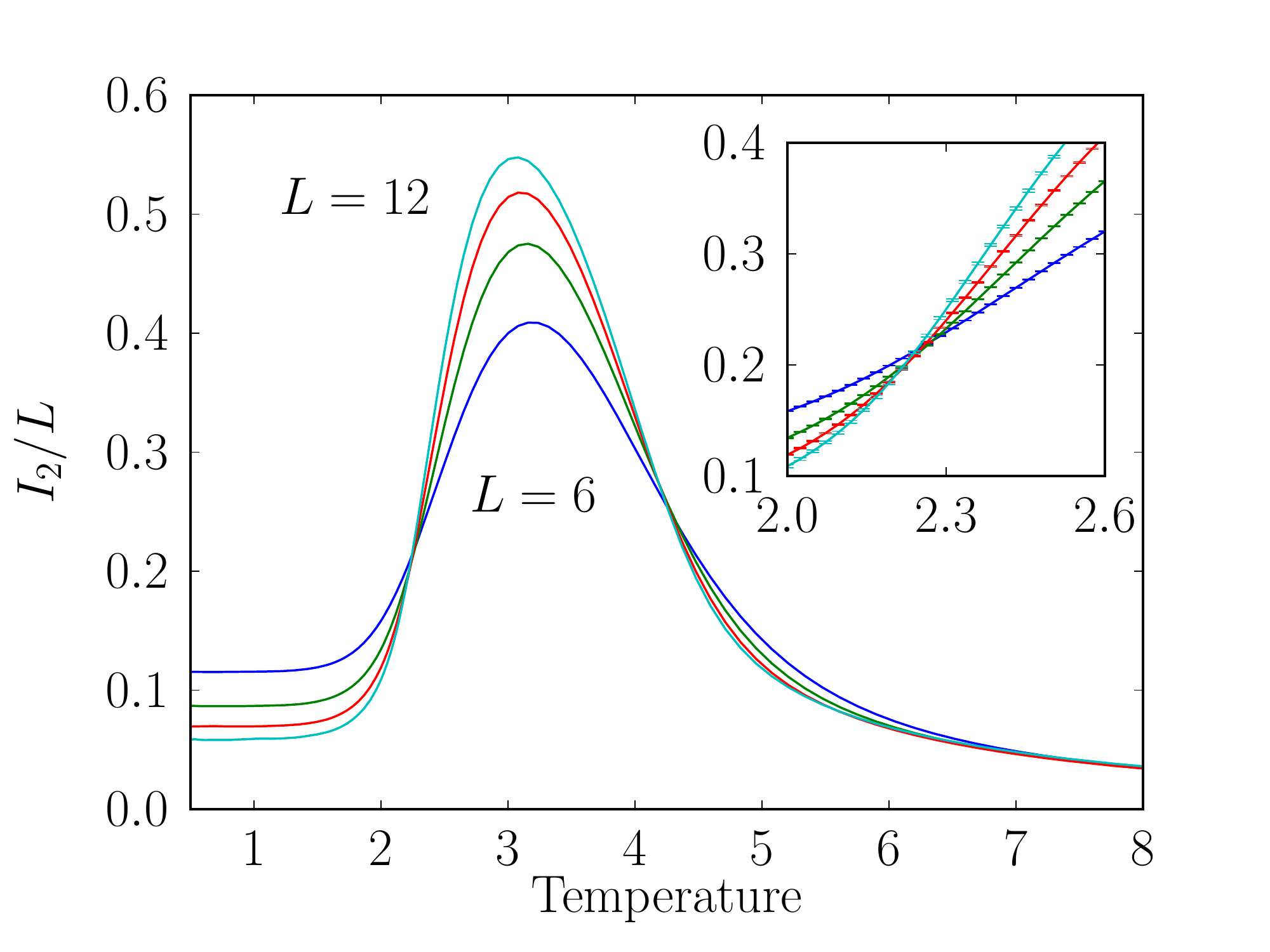}
\caption{The mutual information divided by system size for the 2D transverse field Ising model using the extended broad sampling approach at $\Gamma = J = 1$.
Inset shows a closeup of the lower crossing of mutual information at around $T=2.21$, crossing which we expect to occur at the transition temperature.
Different lines correspond to lattices are of size $6,8,10$ and $12$ (from lowest to highest central peak).
\label{fig:2d_tfim}
}
\end{figure}
To test the extended broad histogram sampling algorithm~\cite{Wessel_MRT} we illustrate results from another Hamiltonian, the transverse field Ising model.
On the 2D square lattice the classical Ising model has a transition temperature of $T_c = 2J/\ln(1+\sqrt{2}) \approx 2.269 J$  from Onsager's exact result \cite{onsager_ising}.
When we add a small transverse field we expect that the transition temperature will be slightly lowered, and as the critical point of this model for the two dimensional system is $\Gamma_c = 3.044 J$ we are well within the Ising like phase at $\Gamma = J$ \cite{blote_tfim}.
Figure~\ref{fig:2d_tfim} shows the mutual information curves for this model for a small set of sizes, showing crossings around $T_c \approx 2.21$, slightly below the non-perturbed result, as expected.
The inset shows a very tight error bound for the data points when using the extended broad histogram technique.

\begin{figure}
\includegraphics[width=0.85\columnwidth]{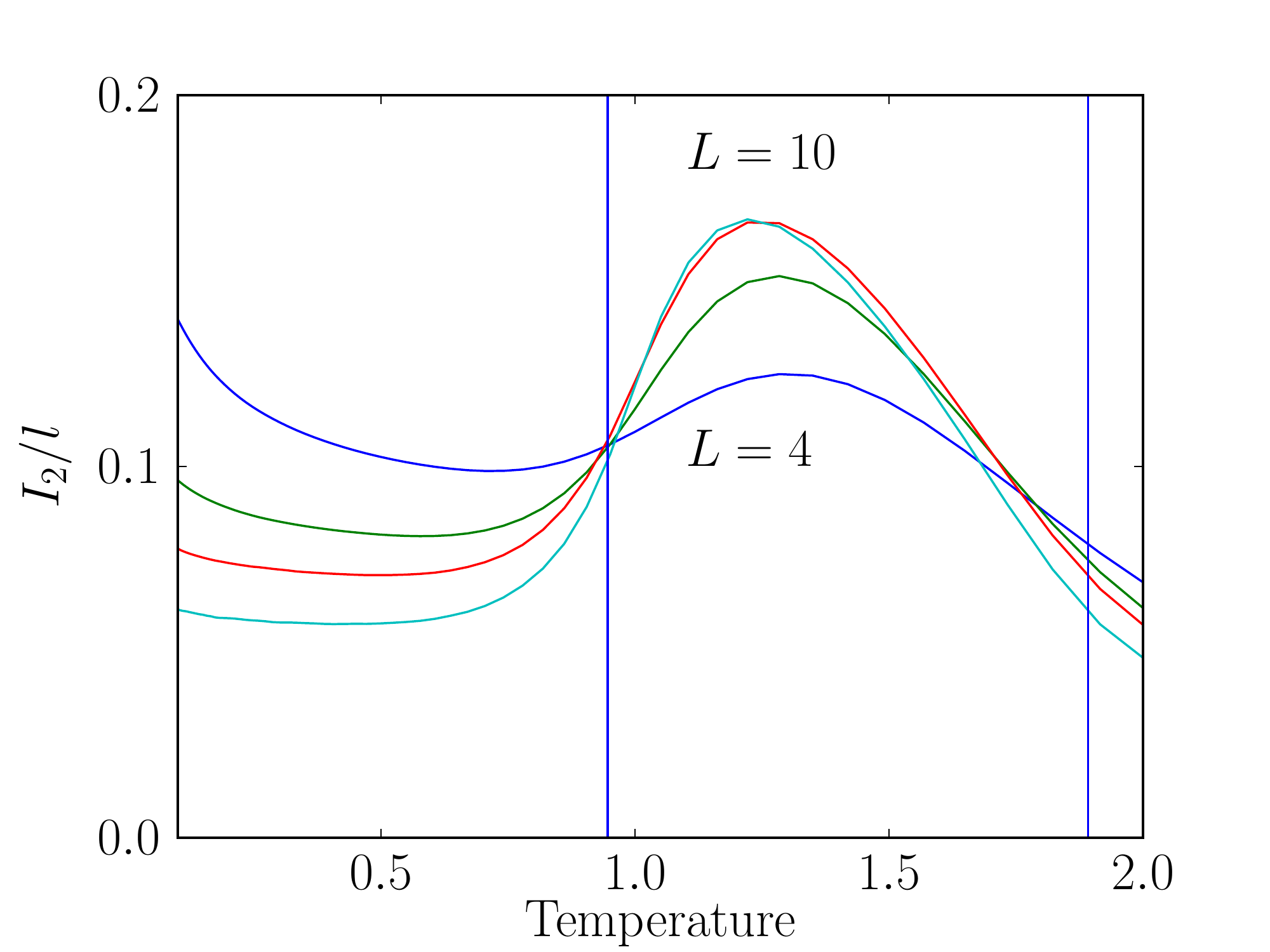}
\caption{The energy per spin and mutual information divided by surface area (still denoted $l$) for the 3D Heisenberg model.
The crossings of the mutual information very close to $T_c$ and $2T_c$, marked with vertical lines.
Different lines correspond to lattices are of size $4,6,8$ and $10$ (from lowest to highest central peak).
\label{fig:3d_heis}
}
\end{figure}

Finally, we examine the performance of this method in higher dimensions using the 3D isotropic Heisenberg model.
Unlike the 2D case, the 3D Heisenberg model does have a finite temperature phase transition, occurring for $\Delta=1$ at $T_c = 0.946$ \cite{sandvik_3dheis}.
In this case, the regions A and B used to calculate the Renyi entanglement entropy consisted of a periodic {\it plane} spanning the 3D simulation
cell, cutting it into two halves.
MI results are illustrated in Fig.~\ref{fig:3d_heis} for a largest size of $10\times 10\times 10$.  
In Fig.~\ref{fig:3d_heis}, the MI per surface area of this plane exhibits crossings approach the critical temperature within error bars.  Interestingly, finite size effects are magnified near $2T_c$, when compared to the 2D transition studied above.  It would be interesting to develop a full scaling theory for the 3D Heisenberg transition to examine these effects in the MI, similar to what was done previously in 2D \cite{roger_MI}.

\section{Discussion}

Although Wang-Landau simulations in themselves are not new \cite{WL}, the use of such non-canonical sampling techniques in quantum Monte Carlo algorithms is still being explored.
In general, techniques like Wang-Landau tend to suffer from slower convergence for a specific parameter (i.e.~the energy at a specific temperature) when compared to importance sampling techniques, but the trade-off is that Wang-Landau generates this result for all temperatures, something that would require many more simulations using importance sampling.
In the end, the decision of the type of sampling to use should depend on the nature of the most important observable.

Wang-Landau can also be extended to measure any quantity that could be measured using traditional SSE.
Instead of just simply having $g(n)$, we now associate any number of other observables (like magnetization) with each $g(n)$.
By looking at the average value sampled for a given $n$, when we re-weight all the $g(n)$ we can calculate these quantities for any temperature.
One difference between Wang-Landau and importance sampling here is that because all energies are sampled, the simulation does not tend to get stuck in local configurations.
A system with a symmetry broken ground state, like the 2D Ising model, will sample all ground states with higher frequency than most importance sampling methods.
This requires some modified sampling styles similar to that needed with parallel tempering, but it is not difficult to implement.

We have demonstrated that the Wang-Landau technique can be successfully used along with a modified SSE formalism \cite{roger_multisheet_sse} to generate the Renyi entanglement entropies, and associated mutual information, over a wide range of temperature.
We confirm that extending simulations to use the modified $1/t$ schedule ensures that any simulation error is systematically reduced by continued sampling.
Improvements using optimized ensembles allows for a similar analysis once the density of states is extracted.

Such broad histogram techniques can also be extended to work in other parameters of the Hamiltonian, such as the size of the entangled region A in the replica trick.
A Wang-Landau in this parameter would allow the calculation of the entanglement entropy at a fixed temperature without integration.
In fact, the implementation by Humeniuk \cite{roscilde_qmc} is identical to the implementation of such a Wang-Landau in region size, with only minor changes.
As the accumulation of error is still a concern, it is not trivially clear that integrating in energy or integrating in region size leads to a smaller error for all cases.
This should be explored in the future.

Using our Wang-Landau method, we have obtained mutual information results for the 2D anisotropic Heisenberg model, reproducing the crossings at the critical temperature \cite{roger_MI}.
For the isotropic case, no critical scaling of the mutual information is observed, consistent with the lack of a finite-temperature phase transition in the model.
We have also implemented a QMC for the 2D transverse field Ising model, where the mutual information again shows evidence of the expected finite temperature phase transition.
Finally, we examine the mutual information in a three dimensional Heisenberg model, and show that crossings occur at $T_c$ and approximately $2T_c$.

The authors thank P.-N. Roy, R. Singh, S. Trebst, and A. Sandvik for valuable discussions, and the Quantum Condensed Matter Visitor's Program at Boston University for hospitality during a visit.
This work was supported by NSERC of Canada and Vanier CGS.
Simulations were performed on the computing resources of SHARCNET.

\bibliography{stephen}{}

\end{document}